\documentstyle[prd,tighten,aps]{revtex}
\begin{document}
\draft
\twocolumn[\hsize\textwidth\columnwidth\hsize\csname
@twocolumnfalse\endcsname  

\title{ 2-D Gravity and the Extended Formalism }
\author{Fernando P. Devecchi}
\address{Depto. de F\'{\i}sica, Universidade Federal do Paran\'a,
cx. postal 19091, cep 81531.990, Curitiba-PR, Brazil. }
\maketitle

\begin{abstract}
The role  of $SL(2,I\!\!R)$ symmetry in two-dimensional gravity is
investigated in the context of the extended Hamiltonian formalism. 
 Using our results we clarify   previous works
 on the subject.
\end{abstract}
 

\vskip2pc]   

\section{introduction} 

The analysis of symmetries and quantization of the induced two-dimensional
gravity model (2-D gravity) proposed by Polyakov\cite{Pol} has received
 the attention of several authors \cite{Abd}\cite{Gho}\cite{Arm} . In the
 original work a ``residual" symmetry appeared when the model was
 studied in the light-cone gauge\cite{Pol}; the generators satisfying
 an $SL(2,I\!\!R)$ algebra. In following papers this feature was approached
 with a variety of techniques. An important idea arising from these works
 is that to understand the $SL(2,I\!\!R)$ symmetry a gauge-independent
  analysis
 is fundamental,   trying to confirm that this invariance is something
 basic in 2-D gravity. The first gauge invariant formulation\cite{Abd}
 arrived at the conclusion that the $SL(2,I\!\!R)$ algebra, realised
  by generalised
 currents,  made  sense  as a symmetry only in the light-cone gauge. In\cite
{Gho}
 the problem was studied in the context of improper gauge
 transformations, but the results were not completely conclusive. Finally,
 in\cite{Arm}, working with the canonical Hamiltonian formalism, it was
  concluded
  that the $SL(2,I\!\!R)$ symmetry arised on the classical level only when
  the $x^+$ coordinate was taken as time.
  
                    In this work we propose to clarify this problem working
with the {\it extended} Hamiltonian formalism. Adopting  strictly this
 technique, we
show that it is possible to understand the role and origin of $SL(2,I\!\!R)$
 symmetry
 when we impose a gauge fixing.  Instead, what it is usually found in the
literature
  is the  {\it direct} injection of the gauge conditions in the 
 original
action, an approach that makes impossible to elucidate
 the role of any
residual symmetry. 
 
  The paper is organized as follows. In Section II we make a short
  description and comments of the approaches  found in the literature.
  In section III we present the fundamental ideas behind the extended
  Hamiltonian formulation and how it works with the problem of residual
   symmetries. Section IV show our results when we apply the
   method to the  induced gravity model. Finally, in Section V we present
   our conclusions.

\section{Gauge invariant  and reduced phase-space approaches} 

 This section is devoted to a brief description of Polyakov's induced gravity
 model and of previous works related to the  $SL(2,I\!\!R)$ symmetry,
distiguishing basically two approaches: the gauge invariant approach and the 
reduced phase-space approach. 
  
  The two-dimensional induced gravity model \cite{Pol} has a rich gauge
structure. In order to take advantage of this feature (to find the physical
solutions or to understand the role of $SL(2,I\!\!R)$ invariance,
 for instance)
  it is important to have first a consistent gauge invariant
formulation.

            The basic ideas in the gauge invariant analysis begin to get
	     shape
 when we manage to write the action as a local functional (introducing an
auxiliary  scalar field $\phi (x)$)\cite{Abd}

\begin{equation}
S=\int d^2x \sqrt{-g} \left(-\phi \Box \phi - \alpha R\phi +\alpha ^2\beta
\right) \,\, ,\label{2}
\end{equation}
where $R$ is the two-dimensional scalar curvature and

\begin{equation}
\alpha ^2=8k-\frac{1}{12\pi}\,\,\,\,\, \beta =-\mu ^2\left(\frac{2k}{\alpha
^2}\right)\,\, ,\label{3}
\end{equation}
$k$ being a function of the central charge of the original model (gravity
coupled to matter) and $\mu$ is the cosmological constant.

Starting with (\ref{2}) it is possible to construct the classical Hamiltonian
 formulation.
The diffeomophism invariance present in this  model imply into well-known
 expressions
for the canonical Hamiltonian density $H_c$  and primary (first class)
constraints
$\pi^{00}$ and $\pi ^{01}$, $\pi  ^{\mu \nu }$ being the momenta canocally
 conjugated to the metric
components $ g_{\mu \nu }$

\begin{equation}
H_c=-\frac{\sqrt {-g}}{g_{11}} \phi _1+ \frac{g_{01}}{g_{11}}           
 \phi _2\,\, ,\label{4}
\end{equation}

\begin{mathletters}
\label{5}                 
\begin{eqnarray}
\pi ^{00}\approx &0&, \label{mlett:1}  \\
\pi ^{01}\approx &0&, \label{mlett:2}
\end{eqnarray}
\end{mathletters}
where $\phi _1$ and $\phi _2$  are secondary (first class)
 constraints,
that follow from the time consistency of (\ref{5})

\begin{mathletters}
\label{6}
\begin{eqnarray}
  \phi _1 =\frac{1}{2} \left(\phi ^{'2}-\frac{4}{\alpha
^2}(g_{11}
\pi ^{11})^2-
\frac{4}{\alpha }(g_{11}\pi ^{11})\pi \right.\nonumber \\ \left. -\alpha 
\frac {g_{11}'}{
g_{11}}+2\alpha \phi '' +\alpha ^2\beta g_{11}\right) \, ,\label{mlett:1}\\
\phi_2=\pi \phi ' -2g_{11}\pi ^{11'}-\pi ^{11}g_{11'}\,\,
 \label{mlett:2},       
\end{eqnarray}
\end{mathletters}
 $\pi$ is the momentum canonically conjugated to the scalar field
$\phi(x)$.
The set of first class constraints showed above represent, as usual,
 the Hamiltonian
generators of diffeomorphism invariance. An important
feature here is that it is possible in this
context to obtain some information about  the residual $SL(2,I\!\!R)$
. In[3],  Abdalla et al. proposed   the construction of a
 generalization of the light-cone gauge
 currents ($J(x)$)
\begin{mathletters}
\label{60}
\begin{eqnarray}
J^+&=&\frac{1}{g_{11}}\left( \phi _2-\phi _1\right)+\frac{1}{2} \alpha ^2
\beta, \label{mlett:1}\\
J^0&=&j^0-x^-J^+, \label{mlett:2}\\
j^0&=&\sqrt 2 \left[ g_{11}\left(\pi^{11}+\frac {\alpha}{2}\frac{\phi
'}{g_{11}}\right)\right. \nonumber \\ &+&\left.\frac{\alpha}{2}\left(\pi -
\frac{\alpha }{2}\frac{
g_{11}'}{g_{11}}-\phi '\right)\right]\, , \\
J^-&=&j^- -2x^-J^0-(x^-)^2J^+,\label{mlett:3}\\
j^-&=&\alpha ^2\left(g_{11}+1\right),
\end{eqnarray}
 \end{mathletters}
that satisfy, as their light-cone partners,  the well-known $SL(2,I\!\!R)$ 
   algebra

\begin{equation}
\{J^a(x),J^b(y)\}=-2\sqrt 2 \epsilon ^{abc} \eta _{cd} J^d (x)
\delta (x-y)\,\, .\label{7}
\end{equation}

The crucial point here  is that these generalised currents  represent symmetry
generators {\it only} in the light-cone gauge, the very $SL(2,I\!\!R)$
 symmetry,  playing no role in  other gauges and therefore loosing
their gauge independent nature.

A different approach to this problem was tried in\cite{Gho} (a reduced
phase-space formulation). The basic
idea was that the $SL(2,I\!\!R)$ symmetry can be interpreted as an inproper
gauge transformation of the action (\ref{2}).
An improper gauge transformation\cite{Giro} appears when the generators of
the local
symmetries $(G)$ need extra terms $(F)$ in order to define unumbiguously
the  field's variations under the action of $\bar G$

\begin{equation}
\bar G(\epsilon)=G(\epsilon)+F(\epsilon)\,\,\, ,
\label {63}
\end{equation}
where $\epsilon$ are the parameters of the gauge transformation.
In the case of Polyakov's induced gravity the $G$'s are simply linear
 combinations
of the first class constraints (\ref{5}) and (\ref{6}). On the other hand,
 $F$ is given
 by

\begin{equation}
F=a_1l_1+a_2l_2+a_3l_3 \,\,\, , \label{70}
\end{equation}
where 

\begin{equation}
\epsilon (x_-,x_+)=a_1(x_+)+x_-a_2(x_+)+(x_-)^2a_3(x_+)\, ,\label{71}
\end{equation}
and
\begin{equation}
l_1=\partial _-, \,\,\, l_2=x ^-\partial _--1, \,\,\, l_3=(x^-)^2\partial
_--2x^-\, . \label{8}
\end{equation}
 The problem is that although the $l_i$'s obey an $SL(2,I\!\!R)$ algebra it
 was  not clear why this quantities
 had to be
associated to the generators of the residual $SL(2,I\!\!R)$ symmetry 
 (as the authors recognize \cite{Gho}).

\section{Residual symmetries and the extended formalism }

In the previous section we have seen that the interpretation for the presence
of the  $SL(2,I\!\!R)$  symmetry in the induced gravity model is full of
 drawbacks.
These problems can be effectively solved if we analize
our model using the Hamiltonian extended formalism\cite{Hen}. This formulation
works with the Dirac's idea that the maximum of information about symmetries
in a gauge theory can be obtained if we consider as a basic ingredient
the so-called extended action

\begin{equation}
S_e=\int \left(p_n\dot q^n-H-\lambda ^a\phi _a-\lambda ^{\alpha} 
\chi _{\alpha}\right)dt \, , \label{11}
\end{equation}
the $\phi $ being the first class constraints and $\chi $ are the second
 class
ones (the $\lambda ^a$ represent their respective Lagrange multipliers).

The formalism gives the following expressions for the canonical gauge
structure
\begin{mathletters}
\label{12}
\begin{eqnarray}
\{\phi _a,\phi _b\}&=&C_{ab}^c\phi _c+T_{ab}^{\alpha \beta} \chi _{\alpha}
\chi _{\beta} \, ,\label{mlett:1} \\
\{\phi _a, \chi _{\alpha}\}&=& C^b _{a\alpha }\phi _b+ C^{\beta}_{a\alpha} 
\chi _{\beta} \, ,\label{mlett:2} \\
\{H, \phi _a\}&=&V^b _a\phi _b+ V^{\alpha \beta}_a\chi _{\alpha }\chi _{\beta}
\, ,\label{mlett:3} \\
\{ H, \chi _{\alpha} \}&=&V^b_{\alpha }\phi _b+ V^{\beta} _{\alpha} 
\chi _{\beta}
\, ,\label{mlett:4}
\end{eqnarray}
\end{mathletters}
the structure functions $C$, $T$ and $V$ being fundamental for our purposes.
The gauge transfomations are  given by 

\begin{equation}
\delta _{\epsilon} F=\epsilon ^a\{F, \phi _a \} \, \, . \label{13}
\end{equation}

In order to the  extended action be invariant under (\ref{13}) the Lagrange
 multipliers
should transform as
\begin{mathletters}
\label{14}
\begin{eqnarray}
\delta \lambda ^a&=&\dot \epsilon ^a+\lambda ^c\epsilon ^bC^{a}_{bc}-
\epsilon ^aV_b, \label{mlett:1} \\
\delta \lambda ^{\alpha}&=&\lambda ^c\epsilon ^bT^{\alpha \beta}_{bc}
\chi _{\beta}-
\epsilon ^bV_b^{\alpha \beta}\chi _{\beta}+\lambda ^{\beta }
\epsilon ^bC^{\alpha }_{b\beta},\label{mlett:2}
 \end{eqnarray} 
 \end{mathletters}

 The important point here is that we can obtain a complete set of symmetries
 of the original action (total action) by simply imposing the 
  gauge fixings
 in the Lagrange multipliers of the secondary constraints ($\lambda ^c$)
 and we can insert them back in the extended action and get the total action.
 More important are the consequeces that this gauge fixing has  on the
 symmetries. Imposing these conditions on (\ref{14}) as
\begin{equation}
\lambda ^c=0 \,\,\,\,\,\,\,
\delta \lambda ^c=0 \, \, ,\label{15} 
\end{equation}
we obtain the symmetries of the total action
\begin{equation}
\delta \Phi(x)=\{\Phi (x), G\},\,\,\,\,\, G= \mu ^a \phi _a  \, , \label{11}
 \end{equation}
where the $ \mu ^a $ must preserve the gauge conditions  (\ref{15}) .  
 
                A very instructive example of this method is the free
Maxwell theory. The extended action reads
\begin{equation}
S_e=\int d^4 \left(\pi ^i\dot A_i+\pi ^0\dot A_0
-H-\lambda ^1 \phi _1-\lambda ^2 \phi _2\right)\, ,\label{17}
\end{equation}
where $\phi _1=\pi^0=0$ is the primary constraint and $\phi _2=\partial _i
\pi ^1=0$ is the secondary (Gauss Law), both are first class.

The generator of the extended action invariances is
\begin{equation}
G=\int d^3x \left( \epsilon ^1 \phi _1+\epsilon ^2 \phi _2\right)\, ,
\label{18}
\end{equation}
with independent gauge parameters $\epsilon$. The commutation relations
between the first class quantities are trivial in this case . The variation
of Lagrange multipliers are\cite {Hen}

\begin{equation}
\delta \lambda ^1=\dot \epsilon ^1\,\,\,\,\,
\delta \lambda ^2=\dot \epsilon ^2-\epsilon ^1\, .\label{19} 
\end{equation}

The usual U(1) invariance of electromagnetism is recovered when we use
the conditions (\ref{15})
\begin{equation}
\delta \lambda ^2=0 \Rightarrow \dot \epsilon ^2=\epsilon ^1 \, .
\end{equation}

\section{The extended formulation for 2-D gravity and The $SL(2,I\!\!R)$ 
  symmetry}

In this section we apply the method described above for the induced gravity
model. The first step is to construct the extended action.  We already
know, from previous sections, the expressions for the canonical Hamiltonian
  and the constraints
structure. So, we  obtain straightforwardly

\begin{equation}
S_e=\int d^2 x\left(\pi ^{00}
\dot g_{00} -2\pi ^{01}
\dot g_{01}
-  \pi ^{11}
\dot g_{11}-
H_c -\lambda ^i\phi _i\right) \,\, .
\label{73}
\end{equation}

 Following (\ref{12}) we see that each of the first class constraints
 , $\phi _i\,\, i=1,..,4$,
 will generate an independent local gauge transformation
\begin{equation}
G=\int dx \left(\epsilon _i\phi ^i\right)\, , \label{74}
\end{equation}
that leave the extended action (\ref{17}) invariant, given the correct
transformations for the Lagrange multipliers. To obtain these transformations
we must use first (\ref{12}) to find the structure functions
 $C^c_{ab}$ and $V^b_a$ (the others being zero 
 because in this case we  have just first class constraints). After some
  manipulations
we find explicitly

\begin{equation}
\{\phi _3(x),\phi _4 (y)\}=\int dz C^3_{34}(x,y,z) \phi (z)
\, ,\end{equation}

where
\begin{equation}
C^3_{34}(x,y,z)=\delta (z-y)\delta ' (z-y)+\delta (z-y) \delta '(x-z)
\, ,\label{20}
\end{equation}
the other non-zero $C$'s are
$C^4_{33}(x,y,z)=C^4_{44}(x,y,z)$ with identical expressions to (\ref{20}).

 We also have for the $V$'s the following expressions
\begin{mathletters}
\begin{eqnarray}
\{H,\phi _3(x)\}&=&\int dz dy V^3_3(x,y,z)\phi _3(z)\nonumber \\ +
\int dz dy V^4_3(x,y,z)\phi _4(z)\, , \label{mlett:1}\\
\{H,\phi _4(x)\}&=&\int dz dy V^3_4(x,y,z)\phi _3(z) \nonumber \\ +
\int dz dy V^4_4(x,y,z)\phi _4(z) \, ,\label{mlett:2}
\end{eqnarray}
\end{mathletters}
where
\begin{mathletters}
\label{22}
\begin{eqnarray}
V^3_3(x,y,z)=-\{\frac{\sqrt {-g}
}{g_{11}}(y) ,\phi _3 (x)\}\delta
(x-z)\nonumber \\ 
+C^3_{43}(x,y,z)\frac{g_{01}}{
g_{11}}(y) \, ,\label{mlett:1} \\
V^4_3(x,y,z)= \{\frac{g_{01}}{g_{11}}(y),\phi _3 (x)\}\delta
(x-z)\nonumber \\+
C^4_{33}(x,y,z)\frac{\sqrt {-g}}{g_{11}}(y)\, , \label{mlett:2} \\
V^3_4(x,y,z)=-\{\frac{ \sqrt {-g}}{g_{11}}(y),\phi _4(x)\}\delta (x-z)
\nonumber \\ +C^3_{34}(x,y,z)\frac{ \sqrt {-g}}{g_{11}}(y)\, ,
\label{mlett:3} \\
V^4_4(x,y,z)=\{\frac{ g_{01} }{g_{11}}(y),\phi _4(x)\}\delta
(x-z)\nonumber \\ +
C^4_{44}(x,y,z)\frac{g_{01}}{g_{11}}(y)\, .\label{mlett:4}
\end{eqnarray}
\end{mathletters}

The final step is our most important result. We can obtain the light-cone
formulation going to the total action formulation imposing the conditions
(\ref{15}), that read in our model
\begin{equation}
\label{23}
\lambda ^3=0=\lambda ^4\,\,\,\,\, \delta \lambda ^3=0=
\delta \lambda ^4\, ,
\end{equation}
into the secondary constraint's  Lagrangian multipliers variations

\begin{eqnarray}
\delta \lambda ^3(x)=\dot \epsilon ^3(x)\nonumber \\
+\int dy\left([\lambda^3(x)\epsilon
^4(x)+\lambda^4(x)\epsilon
^3(x)] \right. \nonumber \\ \left. +[\lambda^3(y)\epsilon
^4(y)+\lambda^4(y)\epsilon
^3(y)]\right)\delta '(x-y)\nonumber \\   +\int dz dy
\left(V^3_3(x,y,z)\epsilon ^3(y)+
V^3_4(x,y,z)\epsilon ^4(y)\right)\, ,
\end{eqnarray}

\begin{eqnarray}
\delta \lambda ^4(x)=\dot \epsilon ^4(x)\nonumber \\
+\int dy\left([\lambda^3(x)\epsilon
^4(x)+\lambda^4(x)\epsilon
^3(x)]\nonumber \right.\\  \left.+[\lambda^3(y)\epsilon
^4(y)+\lambda^4(y)\epsilon
^3(y)]\right)\delta '(x-y)\nonumber 
\\  +\int dz dy\left(V^4_3(x,y,z)\epsilon ^3(y)+
V^4_4(x,y,z)\epsilon ^4(y)\right)\, .
\end{eqnarray}

These relationships define restrictions on the gauge parameters
  $\epsilon(x)$ and the basic fields.
   Using these expressions in the original
 gauge transformations (diffeomophism invariance) for the basic fields
 $g_{\mu \nu}(x)$ and $\phi (x)$\cite{Abd}
we obtain, after a tedious  calculation ($g_{++}=\frac{1}{2}(g_{00}+2g_{01}
+g_{11})$ , $\epsilon^{\pm }= \frac{1}{\sqrt 2}(\epsilon ^3 \pm \epsilon ^4
 )$)

\begin{mathletters}
\label{26}
\begin{eqnarray}
\delta \phi &=&\epsilon ^-\partial _-\phi-\alpha \epsilon ^-\, ,\label{mlett:1
}
\\
\delta g_{++}&=&\epsilon ^-\partial _-g_{++}-g_{++}\partial _-\epsilon ^-
-\partial _+ \epsilon ^-\, ,\label{mlett:2}
\end{eqnarray}
\end{mathletters}

which are exactly the  transformations generated by the so-called
 $SL(2,I\!\!R)$ currents (\ref{7}) in the light-cone gauge.

We also  verify that when we substitute these conditions  on the extended
action we obtain, following the method's prescription, the light-cone gauge
  action

\begin{equation}
S=\int dx \left(\pi _{\phi}\dot \phi +\pi _g\dot g_{++}-L_{lc}\right)
\,\, , \end{equation}
where $L_{lc}$ is the light-cone gauge Lagrangian

\begin{equation}
L_{lc}=\partial _+\phi \partial _-\phi+g_{++}(\partial _-\phi )^2-\alpha 
\partial _-g_{++}\partial _-\phi \,\, .
\end{equation}

As we see the  expressions (\ref{26})  are obtained as a by-product
of the extended formulation making clear the role of the $SL(2,I\!\!R)$ 
symmetry in the induced gravity model and the relation with  the
 light-cone formulation
as a whole.
\section{Conclusions}

               In this work we have clarified the role of classical
	       $SL(2,I\!\!R)$   symmetry using the extended Hamiltonian
 formalism.
This formulation leave intact the separation between physical and
spurius degrees of freedom, making the process of gauge fixing in
induced gravity unambigous. In  early works mentioned instead, the
light-cone gauge conditons were injected directly on the original action
making obscure the origin and role of the $SL(2,I\!\!R)$  symmetry as a
 residual
symmetry. On the other hand, the gauge independent formulations also
mentioned here were inconclusive about this issue.

{\bf Acknowledgements} 

The author would like to Thank Prof. Marc Henneaux for suggesting the approach
used on the calculations.

\end{document}